\documentclass[journal]{IEEEtran}
\usepackage{amsmath,amsfonts}
\usepackage{algorithmic}
\usepackage{algorithm}
\usepackage{array}
\usepackage[caption=false,font=normalsize,labelfont=sf,textfont=sf]{subfig}
\usepackage{textcomp}
\usepackage{stfloats}
\usepackage{url}
\usepackage{verbatim}
\usepackage{graphicx}
\usepackage{cite}
\usepackage{amssymb}
\begin{document}

\title{Joint Channel Estimation and Turbo Equalization of Single-Carrier Systems over Time-Varying Channels}

\author{Yifan Wang,~Minhao Zhang,~Xingbin Tu,~Zhipeng Li,~Fengzhong Qu,~\IEEEmembership{Senior Member,~IEEE},~Yan Wei
\thanks{Yifan Wang, Minhao Zhang, Xingbin Tu and Zhipeng Li are with the Key Laboratory of Ocean Observation-Imaging 
Testbed of Zhejiang Province, Zhejiang University, Zhoushan 316021, China, and also with the Engineering 
Research Center of Oceanic Sensing Technology and Equipment, Ministry of Education, Zhoushan 316021, 
China (e-mail: wyfssgn@zju.edu.cn; z\_mh@zju.edu.cn; xbtu@zju.edu.cn; joeli@zju.edu.cn).}
\thanks{Fengzhong Qu and Yan Wei are with the Key Laboratory of Ocean Observation-Imaging 
Testbed of Zhejiang Province, Zhejiang University, Zhoushan 316021, China, also with the Engineering 
Research Center of Oceanic Sensing Technology and Equipment, Ministry of Education Zhoushan 316021, China, 
and also with Hainan Institute of Zhejiang University, Sanya 572025, China (e-mail: jimqufz@zju.edu.cn; redwine447@zju.edu.cn).}
}

\markboth{May 16, 2023}
{Shell \MakeLowercase{\textit{et al.}}: A Sample Article Using IEEEtran.cls for IEEE Journals}


\maketitle

\begin{abstract}
Block transmission systems have been proven successful over frequency-selective channels. 
For time-varying channel such as in high-speed mobile communication and underwater communication, existing equalizers assume that channels over different data frames are independent. 
However, the real-world channels over different data frames are correlated, thereby indicating potentials for performance improvement. 
In this paper, we propose a joint channel estimation and equalization/decoding algorithm for a single-carrier system that exploits temporal correlations of channel between transmitted data frames. 
Leveraging the concept of dynamic compressive sensing, our method can utilize the information of several data frames to achieve better performance. 
The information not only passes between the channel and symbol, but also the channels over different data frames. 
Numerical simulations using an extensively validated underwater acoustic model 
with a time-varying channel establish that the proposed algorithm outperforms the former bilinear generalized approximate message passing equalizer and 
classic minimum mean square error turbo equalizer in bit error rate and channel estimation normalized mean square error. 
The algorithm idea we present can also find applications in other bilinear multiple measurements vector compressive sensing problems.
\end{abstract}


\section{Introduction}
\IEEEPARstart{T}{he} time-varying nature of the channel has always been a challenging problem in 
communication systems. To mitigate this issue, conventional approach is to partition a large 
data frame into several smaller data frames and consider the channel to be static while transmitting each 
small data frame\cite{kaleh1995channel}. This approach has been widely adopted in many modulation systems, including 
single-carrier (SC) and 
orthogonal frequency division multiplexing (OFDM) system for its effective performance.\par
In most of the systems that use block transmission, the channels of 
different time frames are supposed to be independent\cite{zhou2014ofdm}. 
However, in practice, since the channel is generated by the physical surroundings, 
there must be some relationships between the channels across the time domain that can be exploited to enhance the systems' performance. 
\subsection{Joint channel estimation and equalization}
Equalization of channels has always been a significant topic in wireless communication. In general, there are two
main techniques in wireless communication that address this issue. 
One is based on channel estimation (CE) which assumes that the receiver is aware of the channel, e.g., 
minimum mean square error (MMSE) equalization \cite{tuchler2002minimum} and 
frequency domain equalization with generalized approximate message passing (FD-GAMP) \cite{guo2013iterative}. 
The alternative technique employs directive equalizers that use adaptive algorithms, such as least mean square (LMS), recursive 
least squares (RLS) or their variants, and principles such as minimum symbol error rate (MSER) rather than MMSE principle, 
for example, adaptive minimal symbol error rate (AMSER) \cite{gong2013normalized} and 
improved proportionate normalized minimum symbol error rate equalizer (IPNMSER) \cite{xi2020sparsity}.\par

CE-based equalizers provide excellent performance. However, in low signal-to-noise ratio (SNR) scenarios, 
the reliability of channel estimated by the pilot degrades significantly, and the symbol detection 
in the equalization phase is affected as well. 
One possible way to calibrate the channel estimation result would be to use the symbol that comes out of the turbo equalizer \cite{7404487}. 
However, the channel estimation is a hard, rather than soft estimation, 
which could result in the error being magnified throughout the process. 
Regarding adaptive equalizer, they need a long pilot to converge 
their tap coefficient, and equalizers that use decision feedback structures might experience performance degradation due to 
error propagation in low SNR \cite{966079}.\par

The joint channel estimation and equalization/decoding method is an innovative equalizer. 
One recent work that leverages the bilinear generalized approximate message passing (BiGAMP) \cite{7426735} designs a joint equalizer with a full 
soft-input and soft-output (SISO) structure\cite{sun2018joint}. The term ``full'' here means that not only is the symbol a soft-input and soft-output as in 
a normal turbo equalizer, \IEEEpubidadjcol
but also the channel. The estimation results of the channel have their expectations and variances, which means every tap has 
its own degree of confidence. In algorithms based on message passing, this prevents errors from being amplified by some erroneously estimated channel 
taps similar to what the low-density parity check (LDPC) decoder does to the input log-likelihood ratios (LLRs) with small values \cite{910572}. 
Therefore, it can outperform traditional CE-based algorithms in both the bit error rate (BER) and normalized mean square error (NMSE) of the channel. 
It has shown significant advantage in burst data transmission scenarios\cite{qin2023joint} compared to adaptive equalizers and CE-based equalizers. 
However, this algorithm fails to address the time-varying nature of the channel.
\subsection{Dynamic Compressive Sensing}
To take advantage of the time-varying nature of the channel, we adopt the concept of dynamic compressive sensing. 
In reality signals are usually correlated in the time domain, such as video reconstruction\cite{6868277} and time-varying 
channel estimation \cite{4445731}. 
A variety of compressive sensing (CS) algorithms have been proposed that leverage measurements over different time frames. 
One group of approaches is based on convex relaxation, such as group-fused Lasso, dynamic Lasso \cite{5201168} and modified-CS algorithm
\cite{dai2011gaussian}. The alternative group is based on Bayesian theory, such as T-MSBL \cite{zhang2011sparse} and DCS-AMP \cite{ziniel2013dynamic}.
SBL-based algorithms outperform the others in estimating signals that are not sparse, 
or whose sensing matrix columns have significant correlation \cite{wipf2011sparse}, while AMP-based techniques converge quickly 
and have minimal computational complexity. 
In our communication problem model, the 
columns of the sensing matrix are not highly correlated, so DSC-AMP will not suffer significant performance degradation. 
Moreover, it is also also belongs to the family of approximate message passing algorithm, similar to BiGAMP, 
making it easier to combine its benefits with the BiGAMP algorithm for solving dynamic compressive sensing (DCS) issues. 
\subsection{Turbo Equalization}
As a fully soft equalizer, the turbo equalization is an indispensable part. 
Turbo equalization is a successful application of Bayesian theory and sum-product algorithm (SPA) \cite{koetter2004turbo}. In turbo 
equalization, both the expectation and variance are provided to the decoder for decoding to calculate extrinsic LLRs. 
Then the LLRs output from the decoder will be regarded as \emph{a prior} probabilities and resent to the equalizer. 
The messaging exchange eventually makes the final LLRs converge close to the \emph{a posteriori} probability, 
although they will never attain true \emph{a posteriori} 
probability. This has been elucidated by belief propagation (BP)\cite{yedidia2003understanding} and SPA\cite{910572}.\par
When the channel impulse response (CIR) is known to the receiver, a trellis-based equalizer\cite{bahl1974optimal}
\cite{li1995optimum} can be used to build a turbo equalizer, but the complexity of the equalizer will increase exponentially with 
the number of CIR taps. In practical scenarios, must equalizers can be modified to a turbo equalizer 
by utilizing LLRs from decoder as \emph{a priori} when estimating the symbols. To our knowledge, the message that passes within 
most turbo equalizers are limited between the equalizer and decoder. 
A BiGAMP-based equalizer can transmit messages between the channel estimator, equalizer and decoder, but only within a single time frame. 
In this paper, we will break this limitation. 
\subsection{Contributions}
In this paper, we propose a new algorithm, referred to as DCS-BiGAMP, 
which introduce the dynamic compressive sensing joint channel estimation and equalization/decoding (DCS-JCED) algorithm inspired by BiGAMP and DCS-AMP. 
Expectation maximization (EM) and LDPC decoder are 
respectively used for updating the \emph{a priori} probability of the channel and symbols. 
We demonstrate that the message in the equalizer can not only be passed between the (SISO) 
equalizer and decoder but also the channel estimator in multiple time domain. 
The proposed equalizer \romannumeral 1) 
exploits the correlation between multiple channels across different time frames, \romannumeral 2) achieves a lower NMSE for channel estimates 
and a better (BER) result compared to the existing equalizer based solely on BiGAMP, and \romannumeral 3) is well suited 
for sparse channels. These claims are supported by numerical simulations with an extensively validated underwater acousti (UWA) channel generated by \cite{qarabaqi2013acoustic}.\par

The scope of this work differs from that of the original equalizer based on BiGAMP\cite{sun2018joint}\cite{qin2023joint} which only
considers time-invariant channels and assumes independence between channels in each frame. 
Here, we utilize the time-varying characteristics across multiple data frames. 

The remainder of the paper is organized as follows. Section \uppercase\expandafter{\romannumeral2} presents the SC 
block transmission system and the channel model. In Section \uppercase\expandafter{\romannumeral3}, we present the 
proposed DCS-BiGAMP joint channel estimation and equalization. We also describe how to integrate it with an LDPC decoder to create a turbo equalizer. 
In Section \uppercase\expandafter{\romannumeral4}, we present benchmark and performance analysis of our equalizers with existing equalizer. Finally, 
we draw the conclusions in Section \uppercase\expandafter{\romannumeral5}. \par
The notations used in this paper are as follows. Boldface uppercase $\mathbf{A}$ denotes a matrix and 
boldface lowercase $\mathbf{a}$ denotes a vector. $a_i$ represents the $i^{th}$ element of $\mathbf{a}$. 
$\mathbf{Z}^{(i)}$ refers to the $i^{th}$ slice along the second dimension of a three-dimensional tensor. 
$\mathbf{I}_M$ denotes the $M \times M$ identity matrix, $\mathbf{0}_{M,N}$ denotes the $M\times N$ 
all-zero matrix, and $\mathbf{1}_M$ denotes the $M \times 1$ all-one vector. $\mathbf{F}_N$ denotes the 
$N\times N$ unitary discrete Fourier transform (DFT) matrix, and $\mathbf{F}_N^{1:L}$ is the matrix 
that includes columns $1$ through $L$ of $\mathbf{F}_N$. $\mathbf{a}[k]$ denotes the vector 
involved in the model of the $k^{th}$ data frame. $( \cdot )^\mathsf{T}$ represents the transpose, 
$( \cdot )^\mathsf{H}$ represents the conjugate transpose, and $(\cdot)^*$ represents the conjugate. 
$\odot$ and $\left\lvert \cdot \right\rvert^{\odot 2}$ represent element-wise multiplication and 
squared absolute value, respectively. $\mu_{a\rightarrow b}$ represents the message 
passed from node $a$ to node $b$. $\text{Ber}(\cdot)$ denotes Bernoulli distribution, $\mathcal{CN}(\cdot)$ denotes 
complex Gaussian distribution. 
\section{Systems Model}
In this section we depict the frame structure of the SC block transmission system and the 
received signal model expressed using tensor and vector multiplication. We also show how to 
use the Gaussian-Markov process to establish a simple time-varying channel model.
\subsection{Frame structure}
In the SC block transmission system, each baseband frame takes the form of
\begin{equation}\label{frame structure}
  \mathbf{x}=[\mathbf{x}_p^\mathsf{T}, \mathbf{x}_c^\mathsf{T}, \mathbf{x}_g^\mathsf{T}]^\mathsf{T},
\end{equation}
where $\mathbf{x}_g=\mathbf{0}_{Ng \times 1}$ is the zero guard sequence with $N_g$ being greater than the
channel length $L$ 
and $\mathbf{x}_{p} \in \mathbb{R}^{Np \times 1}$
is the pilot sequence modulated by binary phase shift keying (BPSK). In this paper, we choose the M-sequence as the pilot sequence. 
$\mathbf{x}_c \in \mathcal{A}^{N_{\text{cd}}
  \times 1}$ is the information sequence, where $\mathcal{A} \triangleq [\alpha_1, \alpha_2 , \dots \alpha_{2^Q}]$
denotes the symbol alphabet, e.g, $Q=2$ indicates QAM modulation.\par
The message sequence $\mathbf{x}_c$ carries $N_b$ information bits $b=[b_1,b_2,\dots,b_{N_b}]^{\mathsf{T}}$ where $\ b_n \in
  \{0, 1\}$ and $n \in [1, N_b]$. These bits
are first passed through an interleaver and then encoded by an encoder to generate a coded bit stream consisting of $N_{\text{code}} = N_b/R$ bits,
where $R$ is the code rate of the encoder. The encoded bits are then divided into $N_{\text{cd}}$ subgroups, i.e. 
$\mathbf{c}=[\mathbf{c}_1^\mathsf{T}, \mathbf{c}_2^\mathsf{T}, \dots, \mathbf{c}_{N_{\text{cd}}}^\mathsf{T}]^\mathsf{T}$, 
where $N_{\text{cd}} = N_{\text{code}}/Q$ and $\mathbf{c}_n = [c_{n,1}, c_{n,2}, \dots, c_{n, Q}]^\mathsf{T}$. Then 
$\mathbf{c}$ is mapped to $\mathbf{x}_c$. \par
The received signal of the $k^{th}$ data frame in the baseband is expressed as
\begin{equation}\label{signal passing channel}
  y_n[k]=\sum_{l=1}^{L}x_{n-l+1}[k]h_{l}[k]+w_n[k], \ n = 1,2,\dots,M, 
\end{equation}
where $h_{l}[k]$ is the $l^{th}$ tap of CIR in the time frame when the $k^{th}$ data frame is transmitted. 
Since we use a quasi-static channel, $h_l$ will not change within one frame. $w_n$ is an additive zero-mean white Gaussian noise
sequence with a variance of $\sigma_w$. We rewrite the 
received signal in matrix form as follows:
\begin{equation}\label{matrix form}
  \begin{aligned}
    \mathbf{y}[k] & = \mathbf{z}[k] + \mathbf{w}[k]                            \\
                & =\sum_{l=1}^{L}h_l[k]\mathbf{Z}^{(l)}\mathbf{x}[k] + \mathbf{w}[k],
  \end{aligned}
\end{equation}
where
\begin{equation}
  \mathbf{Z}^{(l)} =
    \begin{bmatrix}
      \mathbf{0}_{(l-1)\times (M-l+1)} & \mathbf{0}_{(l-1)\times (l-1)}\\
      \mathbf{I}_{M-l+1} & \mathbf{0}_{(M-l+1)\times (l-1)}
    \end{bmatrix},
\end{equation}
and $N_c = M = N_{\text{cd}}+N_p+N_g$.
We also define $\mathbf{z}^{(i,j)}$ as the $j^{th}$ column of $\mathbf{Z}^{(i)}$.
\subsection{Channel model}
The common underwater acoustic time-varying multipath CIR is modeled by the transfer function
\cite{qarabaqi2013acoustic}\cite{2013Statistical}
\begin{equation}
  H(f,t) = \sum_p h_p(t_n)\gamma_p(f,t)e^{-j 2\pi f \tau_p (t)},
\end{equation}
where $\gamma_p(f,t) \ \text{with} \ p=0,1,\dots$ represents the scattering coefficient, 
$\tau_p(t) = \tau_{p0}-a_p t$ represents the time-varying path delay according to the random Doppler scaling factors
$a_p$, and $h_p(t_n)$ represents the path gain. The scattering coefficients are modeled as
complex-valued Gaussian processes, and their statistical properties' correlation in time and frequency are determined
by the variance $\sigma_{\delta p}^2$ and Doppler bandwidth $B_{\delta p}$ of the intra-path delays.\par
The detailed physical model is extremely complicated, 
so we use a simplified model \cite{ziniel2013dynamic} to describe it. The time-varying channel 
can be decomposed into two parts: support and amplitude. Recall that we assume the CIR is static 
in one data frame but time-varying across multiple frames. The channel of the $k^{th}$ time frame $\mathbf{h}[k]$ 
is represented by
\begin{equation}\label{new ch model}
  \mathbf{h}[k]=\mathbf{s}[k]\cdot \boldsymbol{\theta}[k],
\end{equation}
where $\mathbf{s}[k]\in \{0,1\}^{L\times 1}$ represents the support of $\mathbf{h}[k]$ and 
$\mathbf{\theta} \in \mathbb{C}^{L \times 1}$ represents the amplitude of $\mathbf{h}[k]$. 
Both the support and amplitude change over time.\par

We describe the support using two transition probabilities $p_{01}=p(s_i[k]=0 |s_i[k-1]=1)$ and 
$p_{10}=p(s_i[k]=1|s_i[k-1]=0)$, which determine a first-order Markov chain for $s_{i}[k]$ where $i=1, \cdots, L$, 
and each chain is independent.
The marginal probability of $p(s_i[k]=1)$ is $\lambda$ and we assume that 
the Markov chain is in steady-state. By solving  
\begin{equation}\label{mc}
  \begin{bmatrix}
    p_{00} & p_{01}\\
    p_{10} & p_{11}
  \end{bmatrix}
  \begin{bmatrix}
    1-\lambda \\ \lambda
  \end{bmatrix}
  =
  \begin{bmatrix}
    1-\lambda \\ \lambda
  \end{bmatrix},
\end{equation}
we can obtain $p_{10}=\lambda p_{01}/(1-\lambda)$, which implies $p_{01}$ itself 
can determine how the chain evolves over time. A small $p_{01}$ generates a 
static process, while a greater $p_{01}$ generates a more dynamic process. Note that $1/p_{01}$ indicates the 
average length of a continuous sequence of ones.\par
For the amplitude, we assume that $\mathbf{\lambda}_i[k]$ is 
independent for $i=1,\dots,L$, and we use a Gauss-Markov process to describe it, i.e., 
\begin{equation}\label{GM process}
  \theta_i[k]=(1-\varrho )(\theta_i[k-1]-\zeta)+\varrho  w_i[k] + \zeta,
\end{equation}
where $\zeta \in \mathbb{C}$ is the mean of the process, and $w_i[k] \sim \mathcal{C}(0, \rho)$ 
is a Gaussian driving process. $\varrho  \in [0, 1]$ represents the degree of correlation. If $\varrho =1$, 
the entire process is a simple Gaussian random process, and if $\varrho  = 0$, the process 
becomes a static sequence where all elements are equal to $\theta_i[0]$. Note that (\ref{GM process}) 
implies the conditional and marginal probability of $\theta_i[k]$, i.e., 
\begin{equation}
  p(\theta_i[k]|\theta_i[k-1])=\mathcal{CN}(\theta_i[k]|(1-\varrho )(\theta_i[k-1]-\zeta)+\zeta,
  \varrho ^2 \rho)
\end{equation}
\begin{equation}
  p(\theta_i[k])=\mathcal{CN}(\zeta, \sigma^2),
\end{equation}
where $\sigma^2=\frac{\varrho  \rho}{2-\varrho }$.
\par
We combine the previously described support and amplitude processes to provide the \emph{a priori} probability for CIR of each frame as 
a Bernoulli-Gaussian distribution
\begin{equation}
  p(h_i[k])=(1-\lambda)\delta(h_i[k])+\lambda \mathcal{CN}(\zeta, \sigma^2).
\end{equation}
The parameter $\lambda$ controls the sparsity level of the channel. Although this model  
is straightforward, it captures the most significant features of the channel, namely its time-varying and 
sparsity nature, with parameters denoted by $\mathbf{q}=[p_{01}, \lambda, \zeta, \varrho , \rho]$.
\section{Fully soft Equalizer in Time-varying Channel}
In this section, we present our joint channel estimation and equalization/decoding algorithm based 
on the proposed algorithm DCS-BiGAMP, and show how to perform turbo equalization. 
We also show how to adjust the coefficients of the \emph{a priori} probabilities using the EM tuning algorithm.
\subsection{Message Passing in Bilinear Dynamic compressive sensing}
\label{twoA}
Our fully soft-input and soft-output equalizer estimates both the mean and 
variance of the channel and symbol through the \emph{a posteriori} probability $p(\mathbf{h}[k]|\mathbf{y}[k],\mathbf{q})$ 
and $p(\mathbf{x}[k]|\mathbf{y}[k],\mathbf{d})$. They are simplified to be complex Gaussian distributions in the 
approximate message passing theory, where $\mathbf{d}$ represents the \emph{a priori} probability returned from the decoder. 
The first step involves constructing to construct the total probability formula:
\begin{equation}
  \begin{aligned}
  &p(\bar{\mathbf{y}},\bar{\mathbf{h}},\bar{\mathbf{x}},\bar{\boldsymbol{\theta}},\bar{\mathbf{s}})\\
  &=\prod_{k=1}^K p(\mathbf{y}[k]|\mathbf{z}[k](\mathbf{h}[k], \mathbf{x}[k]))p(\mathbf{h}[k]|\mathbf{s}[k],\mathbf{\theta}[k])
  p(\mathbf{x}[k])\\
  &=\prod_{k=1}^K \left[ \prod_{m=1}^M p(y_m[k]|z_m[k](\mathbf{h}[k], \mathbf{x}[k])) 
  \prod_{i=1}^L p(h_i[k]|s_i[k],\theta_i[k]) \right.\\
  &\times \left. p(s_i[k]|s_i[k-1]) p(\theta_i[k]|\theta_i[k-1] )\prod_{j=1}^{Nc} p(x_j[k]) \right],
  \end{aligned}
\end{equation}
where $\bar{\mathbf{y}}\triangleq\{\mathbf{y}[k]\}_{k=1}^T$, and similar for 
$\bar{\mathbf{h}}, \bar{\mathbf{x}}, \bar{\boldsymbol{\theta}}, \bar{\mathbf{s}}$.\par

\begin{figure}[!t]
\centering
\includegraphics[width=3.5in]{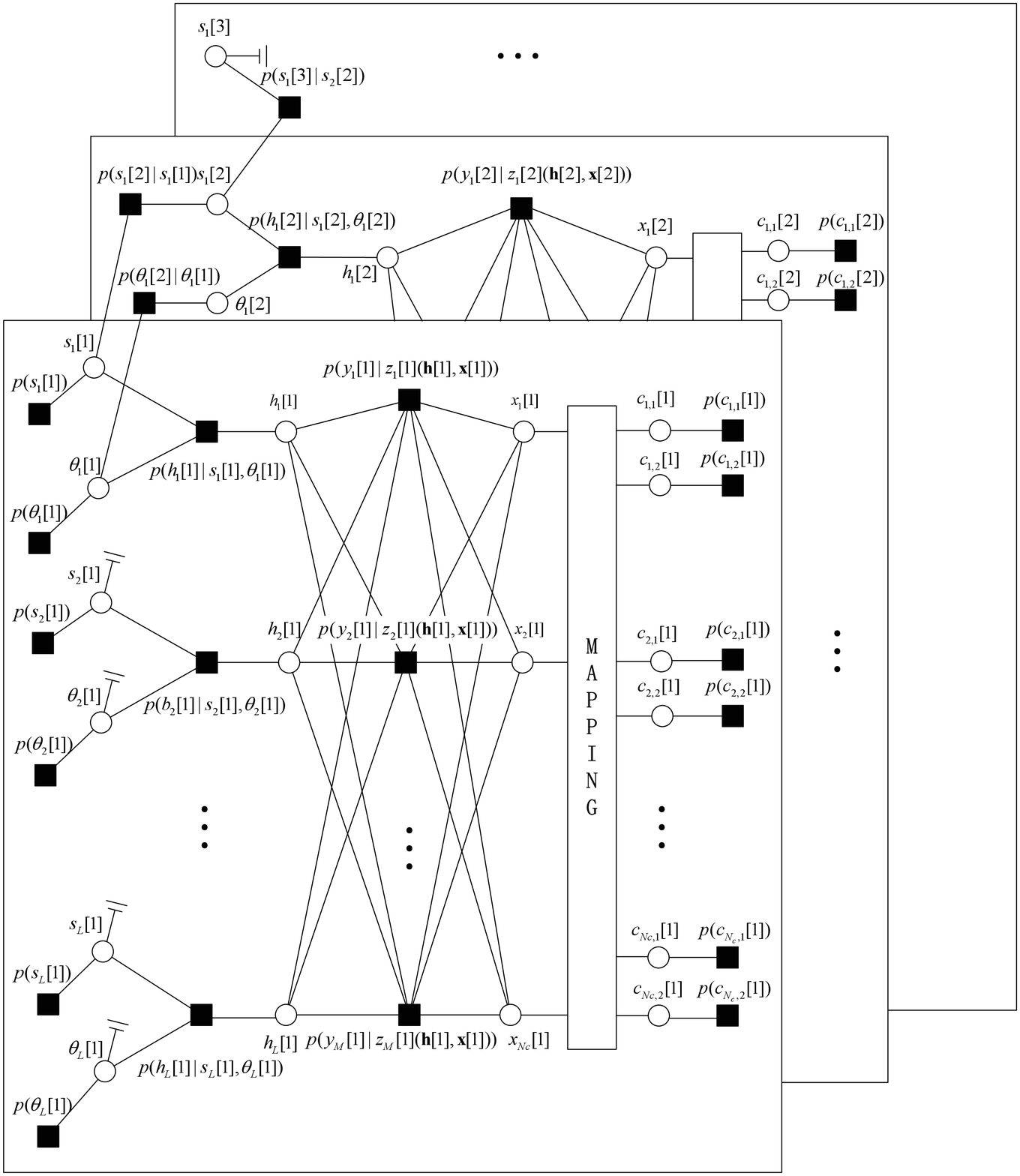}
\caption{Factor graph for DCS-BiGAMP equalizer.}
\label{dsc-bigamp fg}
\end{figure}

Using the total probability formula, we can create a corresponding factor graph, as shown in Fig. \ref{dsc-bigamp fg}.  
Each hollow circle represents a variable node, and each solid block represents a factor node.
A variable node stands for a variable, while a factor nodes stands for a probability density function (PDF).
The edge connecting a variable node and a factor node denotes that the variable is an argument of the PDF.
Here the factor graph for each frame is placed independently in a plane. The correlation of the 
channel is expressed by connecting $\theta_i[k]$ and $\theta_i[k-1]$ with $p(\theta_i[k]|\theta_i[k-1])$, 
and by connecting $s_i[k]$ and $s_i[k-1]$ with $p(s_i[k]|s_i[k-1])$. To make our graph concise, we only include the connection 
between two time frames for $s_i[k]$ and $\theta_i[k]$ when $k = 1,2,3$. 
For a single time frame, we only show connections for $h_i[k]$ when $i=1,2,L$, 
$x_j[k]$, when $j=1,2,N_c$, $p(y_m[k]|z_m[k](\mathbf{h}[1],\mathbf{x}[1]))$ when $m=1,2,M$.
\par
The \emph{a posteriori} probabilities of $p(\mathbf{h}[k]|\mathbf{y}[k],\mathbf{q})$ 
and $p(\mathbf{x}[k]|\mathbf{y}[k],\mathbf{d}[k])$ are computed using the BiGAMP algorithm, which is a 
simplified algorithm based on SPA. Here, $\mathbf{d}[k]$ denotes the \emph{a priori} information of the $k^{th}$ 
data frame feedback from the decoder. 
Although obtaining the true \emph{a posteriori} probabilities 
using SPA requires 
that no loop is allowed in the factor graph\cite{loeliger2004introduction}, which is not achievable in our factor graph in 
Fig. \ref{dsc-bigamp fg}, we can still obtain an approximate \emph{a posteriori} probability. This has 
been successfully demonstrated in many applications of SPA, such as LDPC code \cite{910580}, 
GAMP \cite{6556987}, VAMP \cite{rangan2019vector} and turbo equalization \cite{koetter2004turbo}.\par 
To simplify the expression in writing the message, we need to rename the factor nodes instead 
of referring to them as PDF since they represent corresponding relationships as listed in Table~\ref{tab:rename}, 
before a detailed description of the DCS-BiGAMP algorithm.

\begin{table}[!t]
\caption{Fators and the corresponding PDFs}\label{tab:rename}
\centering
\begin{tabular}{c c}
  \hline
  factor & PDF \\
  \hline
  $g_m^k(\mathbf{h}[k],\mathbf{x}[k])$ & $p(y_m[k]|z_m[k](\mathbf{h}[k], \mathbf{x}[k]))$\\
  $f_i^k(h_i[k],s_i[k],\theta_i[k])$ & $p(h_i[k]|s_i[k],\theta_i[k])$ \\
  $q_i^1(s_i[1])$ & $p(s_i[1])$ \\
  $q_i^k(s_i[k],s_i[k-1])$ & $p(s_i[k]|s_i[k-1])$ \\
  $v_i^1(\theta_i[1])$ & $p(\theta_i[1])$ \\
  $v_i^k(\theta_i[k],\theta_i[k-1])$ & $p(\theta_i[k]|\theta_i[k-1])$ \\
  $d_j^k(x_j[k])$ & $p(x_j[k])$\\
  \hline
\end{tabular}
\end{table}

The DCS-BiGAMP framework is structurally similar to DCS-AMP, which also utilizes 
the SPA algorithm to propagate messages between different temporal frames.
These messages are used to update the ``local prior'' which refers to 
the \emph{a priori} probability of $h_i[k]$: $p(h_i[k]|s_i[k],\theta_i[k])$. The ``local prior'' 
enables the original BiGAMP to function within one time frame. The process can be divided 
into four stages: \emph{into}, \emph{within}, \emph{out} and \emph{across}. In the \emph{into} stage, variable nodes $s_i[k]$ and $\theta_i[k]$ 
gather messages from factor nodes $q_i^k$, $q_i^{k-1}$, $v_i^k$ and $v_i^{k-1}$, and generate new 
messages that are then forwarded to $f_i^{k}$ as the ``local prior'' of $h_i[k]$. In the \emph{within} stage, original 
BiGAMP algorithm is performed to calculate the extrinsic expectation and variance of 
$h_i[k]$ and $x_i[k]$. We denote $\hat{q}_i[k]$ and $\mu^q_i[k]$ as the extrinsic expectation and 
variance of $h_i[k]$, respectively, while the corresponding values for for $x_j[k]$ are $\hat{r}_j[k]$, $\mu^r_j[k]$. 
In the \emph{out} stage, messages from $f_i^k$ to $\theta_i[k]$ and $s_i[k]$ are updated using the 
extrinsic information returned from the \emph{within} stage. Finally, $s_i[k]$ and $\theta_i[k]$ pass the 
messages they gathered from factor nodes $q_i^k$ and $v_i^k$ to update $s_i[k-1]$ and 
$\theta_i[k-1]$. A comprehensive explanation of these four stages will be presented later. 
Note that the framework can run in either serial or parallel mode. 
In serial mode, the \emph{into} stage is performed only 
for $k = 1$, followed by the \emph{within} and \emph{out} stages. The \emph{across} stage only updates $s_i[2]$ and $\theta_i[2]$, 
and then the same flow path is performed for $k = 2,3,\dots,K$. 
In parallel mode, all frames (i.e. $k=1,2, \cdots, K$) run the \emph{into}, \emph{within}, and \emph{out} stages simultaneously
In the \emph{across} stage, the update of $s_i[k]$ and 
$\theta_i[k]$ is still conducted serially. Only a brief explanation of the forward propagation process is presented here. 
The the backward propagation is extremely similar to this, with the exception that the \emph{across} stage need to be modified.\par

\begin{figure}[!h]
  \centering
  \includegraphics[width=3in]{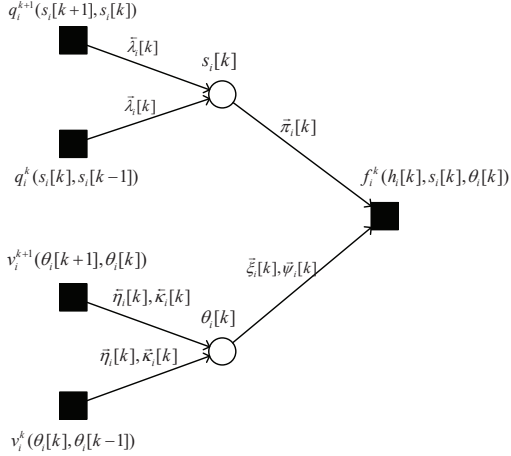}
  \caption{Factor graph for \emph{into} stage.}
  \label{into}
\end{figure}

In the \emph{into} stage as shown in Fig \ref{into}, we need to calculate two messages, $\mu_{s_i[k]\rightarrow f_i^k}$ and 
$\mu_{\theta_i[k]\rightarrow f_i^k}$ using SPA. For message from 
$s_i[k]$ to $f_i^k$, we have
\begin{equation}\label{into ber}
  \begin{aligned}
    \mu_{s_i[k]\rightarrow f_i^k}&=\mu_{q_i^{k+1}\rightarrow s_i[k]}\mu_{q_i^{k}\rightarrow s_i[k]}\\
    &=\text{Ber}(s_i[k]|\overleftarrow{\lambda}_i[k])\text{Ber}(s_i[k]|\overrightarrow{\lambda}_i[k])\\
    &=(1-\overleftarrow{\lambda}_i[k])(1-\overrightarrow{\lambda}_i[k])(1-s_i[k])\\
    &+\overleftarrow{\lambda}_i[k]\overrightarrow{\lambda}_i[k]s_i[k]\\
    &=(1-\overrightarrow{\pi}_i[k])(1-s_i[k])+\overrightarrow{\pi}_i[k]s_i[k],
  \end{aligned}
\end{equation}
where
\begin{equation}\label{into pi}
    \overrightarrow{\pi}_i[k]=\frac{\overleftarrow{\lambda}_i[k]\overrightarrow{\lambda}_i[k]}
    {(1-\overleftarrow{\lambda}_i[k])(1-\overrightarrow{\lambda}_i[k])+\overleftarrow{\lambda}_i[k]\overrightarrow{\lambda}_i[k]}.
\end{equation}\par
Similarly, for message from $\theta_i[k]$ to $f_i^k$, we have
\begin{equation}
  \begin{aligned}
    \mu_{\theta_i[k] \rightarrow f_i^k}&=\mu_{v^{k+1}_i \rightarrow \theta_i[k]}
    \mu_{v^k_i \rightarrow \theta_i[k]}\\
    &=\mathcal{CN}(\theta_i[k]|\overleftarrow{\eta}_i[k], \overleftarrow{\kappa}_i[k])
    \mathcal{CN}(\theta_i[k]|\overrightarrow{\eta}_i[k], \overrightarrow{\kappa}_i[k])\\
    &=\mathcal{CN}(\theta_i[k]|\overrightarrow{\xi}_i[k], \overrightarrow{\psi}_i[k]),
  \end{aligned}
\end{equation}
where
\begin{equation}\label{into psi}
  \overrightarrow{\psi}_i[k]=\left( 
    \frac{1}{\overleftarrow{\kappa}_i[k]}
    + \frac{1}{\overrightarrow{\kappa}_i[k]}
   \right)^{-1}
\end{equation}
\begin{equation}\label{into xi}
  \overrightarrow{\xi}_i[k]=\overrightarrow{\psi}_i[k]
  \left(
  \frac{\overleftarrow{\eta}_i[k]}{\overleftarrow{\kappa}_i[k]}
  +\frac{\overrightarrow{\eta}_i[k]}{\overrightarrow{\kappa}_i[k]}
  \right)
\end{equation}
according to the Gaussian multiplication rule \cite{bromiley2003products}.\par
During the first forward propagation, $\overleftarrow{\lambda}$,$\overleftarrow{\kappa}$
and $\overleftarrow{\eta}$ are not updated by the \emph{across} 
stage of backward propagation, so we initialize them to 0.5, $+\infty$ and 0, respectively. 
In the last frame, i.e., $k = K$, where there are 
no $q^{K+1}$ and $v^{K+1}$, we have $\overrightarrow{\pi}[K] = \overrightarrow{\lambda}[K]$, 
$\overrightarrow{\psi}[K] = \overrightarrow{\kappa}[K]$ and 
$\overrightarrow{\xi}[K] = \overrightarrow{\eta}[K]$. 
Furthermore, $\overrightarrow{\lambda}[1]$ is 
initialized with $\lambda$ and $(\overrightarrow{\eta}_i[1], \overrightarrow{\kappa}_i[1])$ 
is initialized with $(\zeta, \sigma^2)$.
\par

The \emph{within} stage is almost the BiGAMP algorithm itself \cite{7426735}. 
Nonetheless, the original version only provides the \emph{a posteriori} information of the symbol at the end of the iteration, 
whereas a SISO equalizer requires extrinsic information. 
Therefore, we modify the order of the original 
algorithm. The initial input becomes extrinsic variance and expectation instead of \emph{a posteriori} ones, 
and so is the final output. 
In the following, we provide a description of the modified algorithm. 
One can find the frequency domain version in the Appendix. 
The first step of the \emph{within} phase is to calculate the \emph{a posteriori} information using the local prior generated 
in the \emph{into} stage and the LLRs feedback from the decoder. The \emph{a posteriori} probability of the channel is

\begin{equation}
  \begin{aligned}
    &p(h_i[k]|\mathbf{y}[k], \overrightarrow{\pi}_i[k], \overrightarrow{\xi}_i[k], \overrightarrow{\psi}_i[k])\\
    &=(1-\pi_i[k])\delta(h_i[k])+\pi_i[k]\mathcal{CN}(h_i[k]|\gamma_i[k], \nu_i[k] ),
  \end{aligned}
\end{equation}
where
\begin{equation}
  \nu_i[k]=\left( \frac{1}{\mu^q_i[k]}+\frac{1}{\overrightarrow{\psi}_i[k]} \right)^{-1}
\end{equation}
\begin{equation}
  \gamma_i[k]=\nu_i[k]\left(
    \frac{\hat{q}_i[k]}{\mu^q_i[k]}+\frac{\overrightarrow{\xi}_i[k]}{\overrightarrow{\psi}_i[k]}
  \right)
\end{equation}
\begin{equation}
  \pi_i[k] = \frac{1}{
    1+\frac{(1-\overrightarrow{\pi}_i[k])\mathcal{CN}(0|\hat{q}_i[k],\mu_i^q[k])}
    {\overrightarrow{\pi}_i[k]\mathcal{CN}(0|\overrightarrow{\xi}_i[k]-\hat{q}_i[k], \overrightarrow{\psi}_i[k]+\mu_i^q[k])}
  }.
\end{equation}
Hence, the \emph{a posteriori} expectation and variance of $h_i[k]$ are expressed as
\begin{equation}\label{within hhat}
  \hat{h}_i[k] = \pi_i[k]\gamma_i[k]
\end{equation}
\begin{equation}\label{within hvar}
  \mu^h_i[k] = \pi_i[k] \left(
    \gamma_i[k]^2+\nu_i[k]^2-\pi_i[k]\gamma_i[k]^2
  \right).
\end{equation}
Similarly, the \emph{a posteriori} probability of the symbol is 
\begin{equation}
  \begin{aligned}
    &p(x_j[k]=\alpha_n|\mathbf{y}[k],\mathbf{d}[k])\\
    &\propto p_{\text{apr}}(x_j[k]=\alpha_n)
    \mathcal{CN}(x_j[k]|\hat{r}_j[k], \mu^x_j[k]),
  \end{aligned}
\end{equation}
and
\begin{equation}
  p_{\text{apr}}(x_j[k]=\alpha_n)=\prod_q^{Q}p(c_{j,q}=\chi_{n,q}).
\end{equation}
Here, $\chi_{n,q}$ represents the $q^{th}$ bit of the constellation $\alpha_n$, and 
\begin{equation}
  p(c_{j,q})=\frac{\exp\left( (-1)^{c_{j,q}} L_{\text{apr}}(c_{j,q})/2 \right)}
  {\exp\left( L_{\text{apr}}(c_{j,q})/2 \right) + \exp\left( -L_{\text{apr}}(c_{j,q})/2 \right)},
\end{equation}
where $L_{\text{apr}}(c_{j,q})$ denotes the \emph{a priori} LLR of bit $c_{j,q}$ coming from the decoder.
Consequently, the \emph{a posteriori} expectation and variance of $x_j[k]$ can be expressed as
\begin{equation}\label{within xhat}
  \hat{x}_j[k]=\sum_n^{2^Q} p(x_j[k]=\alpha_n|\mathbf{y}[k],\mathbf{d}[k]) \alpha_n
\end{equation}
\begin{equation}\label{within xvar}
  \mu^x_j[k]=\sum_n^{2^Q} p(x_j[k]=\alpha_n|\mathbf{y}[k],\mathbf{d}[k]) |\alpha_n
  -\hat{x}_j[k]|^2.
\end{equation}
\par
The second step of the \emph{within} stage is to calculate the conditional expectation and 
variance of $\mathbf{z}[k]$, namely $p(\mathbf{z}[k]|\mathbf{h}[k], \mathbf{x}[k])$
\begin{equation}\label{within phat}
  \hat{p}_m[k]=\hat{z}_m^{(*,*)}[k]-\hat{s}_m[k] \bar{\nu}_m^p[k]
\end{equation}
\begin{equation}\label{within pvar}
  \nu_m^p[k]=\bar{\nu}_m^p[k]+\sum_{i=1}^{L}\sum_{j=1}^{N_c}\mu^h_i[k]\mu^x_j[k] \left\lvert z_m^{(i,j)}\right\rvert ^2,
\end{equation}
where
\begin{equation}
  \hat{z}_m^{(*,*)}[k]=\sum_{i=1}^L\sum_{j=1}^{N_c}\hat{h}_i[k]\hat{x}_j[k]z_m^{(i,j)}
\end{equation}
\begin{equation}
  \bar{\nu}_m^p[k]=\sum_{j=1}^{N_c} \mu_j^x[k] \left\lvert \hat{z}_m^{(*,j)}[k] \right\rvert^2+
  \sum_{i=1}^L \mu_i^h[k] \left\lvert \hat{z}_m^{(i,*)}[k] \right\rvert^2
\end{equation}
\begin{equation}\label{within shat}
  \hat{z}_m^{(i,*)}[k]=\sum_{j=1}^{N_c}\hat{x}_j[k]z_m^{(i,j)}
\end{equation}
\begin{equation}\label{within svar}
  \hat{z}_m^{(*j)}[k]=\sum_{i=1}^L\hat{h}_i[k]z_m^{(i,j)}.
\end{equation}
We omit the definition of $\hat{s}_m$ which is the first derivative used to simplify the 
message in the inference process of BiGAMP\cite{7426735} here. It will be defined in the next step of 
BiGAMP and used during iteration in this step.\par
In the third step, the \emph{a posteriori} expectation and variance of $\mathbf{z}[k]$ are computed. The Gaussian distribution parameters 
of $p(\mathbf{z}[k]|\mathbf{h}[k], \mathbf{x}[k])$ are provided in the previous step, and 
$p(\mathbf{y}[k]|\mathbf{z}[k]) \sim \mathcal{CN}(\mathbf{y}[k]|\mathbf{z}[k], \sigma_w\mathbf{I})$ in our model, 
so the variance and expectation of $p(\mathbf{z}[k]|\mathbf{y}[k], \mathbf{h}[k], \mathbf{x}[k])$ are 
\begin{equation}\label{within zvar}
  \mu^z_m[k] = \left( 
    \frac{1}{\nu_m^p[k]} + \frac{1}{\sigma_w}
  \right)^{-1}
\end{equation}
\begin{equation}\label{within zhat}
  \hat{z}_m[k]=\mu^z_m[k]\left(
    \frac{y_m[k]}{\sigma_w} + \frac{\hat{p}_m[k]}{\nu_m^p[k]}
  \right).
\end{equation}
In the fourth step, the parameters $\hat{s}_m[k]$ and $\mu^s_m[k]$, which are used for Taylor expansion in BiGAMP, 
are calculated as follows
\begin{equation}\label{within shat}
  \hat{s}_m[k]=\frac{\hat{z}_m[k]-\hat{p}_m[k]}{\nu^p_m[k]}
\end{equation}
\begin{equation}\label{within svar}
  \mu^s_m[k]=\left(1-\frac{\mu^z[k]}{\nu^p_m[k]}\right)\frac{1}{\nu^p_m[k]}.
\end{equation}
The final step of the \emph{within} stage is to estimate the extrinsic parameters of $\mathbf{h}[k]$ and $\mathbf{x}[k]$.
$p_{\text{ext}}(\mathbf{h}[k]) \sim \mathcal{CN}(\mathbf{h}[k]|\mathbf{\hat{q}}[k], \boldsymbol{\mu}^q[k])$ and 
its coefficients are given by
\begin{equation}\label{within qvar}
  \mu^{q}_i[k]=\left(
    \sum_{m=1}^M \nu^s_m[k] \left\lvert \hat{z}_m^{(i,*)}[k] \right\rvert ^2
  \right)^{-1}
\end{equation}
\begin{equation}\label{within qhat}
  \begin{aligned}
    \hat{q}_i[k] &= \hat{h}_i[k] + \mu^q_i[k]\sum_{m=1}^M 
    \hat{s}_m[k]\hat{z}^{(i,*)}_m[k]^*\\
    &-\mu^q_i[k]\hat{h}_i[k]\sum_{m=1}^{M} \mu^s_m[k]\sum_{j=1}^{Nc} 
    \mu^x_j[k] \left\lvert z^{(i,j)}_m \right\rvert ^2.
  \end{aligned}
\end{equation}
$p_{\text{ext}}(\mathbf{x}[k]) \sim \mathcal{CN}(\mathbf{x}[k]|\mathbf{\hat{r}}[k], \boldsymbol{\mu}^r[k])$ and 
its coefficients are computed similarly by
\begin{equation}\label{within rvar}
  \mu^{r}_j[k]=\left(
    \sum_{m=1}^M \nu^s_m[k] \left\lvert \hat{z}_m^{(*,j)}[k] \right\rvert ^2
  \right)^{-1}
\end{equation}
\begin{equation}\label{within rhat}
  \begin{aligned}
    \hat{r}_j[k] &= \hat{x}_j[k] + \mu^x_j[k]\sum_{m=1}^M 
    \hat{s}_m[k]\hat{z}^{(*,j)}_m[k]\\
    &-\mu^r_j[k]\hat{x}_j[k]\sum_{m=1}^{M} \mu^s_m[k]\sum_{i=1}^{L} 
    \mu^h_i[k] \left\lvert z^{(i,j)}_m \right\rvert ^2.
  \end{aligned}
\end{equation}
\par
\begin{figure}[!h]
  \centering
  \includegraphics[width=2.75in]{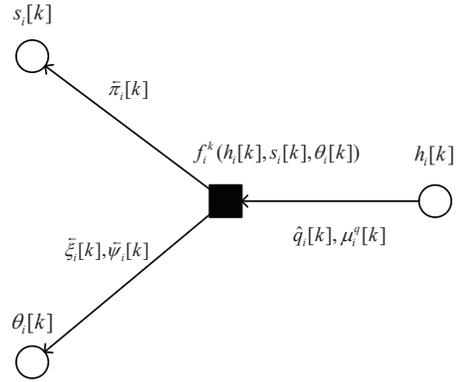}
  \caption{Factor graph for \emph{out} stage.}
  \label{out}
\end{figure}

The entire \emph{within} stage is completed here. The subsequent extrinsic Gaussian distribution parameters of 
$\mathbf{h}$ are used in the \emph{out} stage to transfer the information to the other frame, while the corresponding  
parameters of $\mathbf{x}$ are used to calculate the extrinsic LLRs and send them to the decoder. \par
The \emph{out} stage is illustrated in Fig \ref{out}.
The message from $f_i^k(h_i[k], s_i[k], \theta_i[k])$ to $s_i[k]$ can be computed directly by:
\begin{equation}\label{out message 1}
  \begin{aligned}
    \mu_{f_i^k \rightarrow s_i[k]}&=\int 
    p(h_i[k]|s_i[k],\theta_i[k])\mu_{h_i[k] \rightarrow f_i^k}
    \mu_{\theta_i[k] \rightarrow f_i^k}dh_i[k]d\theta_i[k]\\
    &=\int \delta(h_i[k]-s_i[k]\theta_i[k])
    \mathcal{CN}(\theta_i[k]|\overrightarrow{\xi}_i[k], \overrightarrow{\psi}_i[k])\\
    &\times \mathcal{CN}(h_i[k]|\hat{q}_i[k], \mu^q_i[k])dh_i[k]d\theta_i[k]\\
    &=\int \mathcal{CN}(\theta_i[k]|\overrightarrow{\xi}_i[k],\overrightarrow{\psi}_i[k])\\
    &\times \mathcal{CN}(s_i[k]\theta_i[k]|\hat{q}_i[k], \mu_i^q[k])d\theta_i[k].
  \end{aligned}
\end{equation}
Note that $s_i[k]$ can only be $0$ or $1$, and thus (\ref{out message 1}) can be further simplified as
\begin{equation}
  \begin{aligned}
    \mu_{f_i^k \rightarrow s_i[k]}&=
    \mathcal{CN}(0|\hat{q}_i[k], \mu_i^q[k])(1-s_i[k])\\
    &+\mathcal{CN}(0|\hat{q}_i[k]-\overrightarrow{\xi}_i[k], \mu_i^q[k]+\overrightarrow{\psi}_i[k])s_i[k]\\
    &=(1-\overleftarrow{\pi}_i[k])(1-s_i[k])+\overleftarrow{\pi}_i[k]s_i[k],
  \end{aligned}
\end{equation}
where
\begin{equation}\label{out pi}
  \overleftarrow{\pi}_i[k]=\frac{\mathcal{CN}(m_1,\sigma_1)}{\mathcal{CN}(m_1,\sigma_1)+\mathcal{CN}(0|\hat{q}_i[k], \mu_i^q[k])}
\end{equation}

\begin{equation}
  \begin{cases}
    m_1=\hat{q}_i[k]-\overrightarrow{\xi}_i[k]\\
    \sigma_1=\mu_i^q[k]+\overrightarrow{\psi}_i[k].
  \end{cases}
\end{equation}\par
Updating the message from $f_i^k(h_i[k], s_i[k], \theta_i[k])$ to $\theta_i[k]$ is quite complicated. 
Direct computation will return a Gaussian mixture distribution with two components, 
which is unacceptable in message passing algorithms since Gaussian mixture distributions of multiple elements 
will be generated in later steps during the iteration process. Fortunately, an approach that 
employs Taylor expansion to approximate the Gaussian mixture distribution with a single Gaussian distribution.
$\mathcal{CN}(\theta_i[k]|\overleftarrow{\xi}_[k], \overleftarrow{\psi}_i[k])$ is provided in \cite{6320709}. 
Additionally, an approximation using 
threshold is proposed in \cite{ziniel2013dynamic}. We choose the former for using in DCS-BiGAMP since the 
threshold approximation may lead to numerical instability in numerical simulation. To complete our algorithm, we include 
the approximation flow path 
in Table \ref{tab:taylor}.\par
\begin{table}[!h]
  \caption{Approximation of Gaussian Mixture Distribution}\label{tab:taylor}
  \centering
  \begin{tabular}{|l r|}
    \hline
    Intermediate variable & \\
    \hline
    $\Omega(\overrightarrow{\pi}_i[k])\triangleq\frac{\epsilon^2 \overrightarrow{\pi}_i[k]}{(1-\overrightarrow{\pi}_i[k] + \epsilon^2 \overrightarrow{\pi}_i[k])}
    , \ \epsilon \rightarrow 0$ & (\text{A1}) \\
    $a\triangleq \epsilon^2 (1-\Omega(\overrightarrow{\pi}_i[k]))$ & (\text{A2}) \\
    $\bar{a} \triangleq \Omega(\overrightarrow{\pi}_i[k])$ & (\text{A3}) \\
    $b \triangleq \frac{\epsilon^2}{\mu_i^q[k]} \left\lvert (1-\frac{1}{\epsilon}) \hat{q}_i[k] \right\rvert^2 $ &
    (\text{A4})\\
    $\sigma_r \triangleq -\frac{2\epsilon^2}{\mu_i^q[k]}(1-\frac{1}{\epsilon})\Re \{\hat{q}_i[k]\} $ & (\text{A5})\\
    $\sigma_i \triangleq -\frac{2\epsilon^2}{\mu_i^q[k]}(1-\frac{1}{\epsilon})\Im \{\hat{q}_i[k]\} $ & (\text{A6})\\
    \hline
    Compute outputs & \\
    \hline
    $\overleftarrow{\psi}_i[k]=\frac{(a^2e^{-b}+a\bar{a}+\bar{a}^2e^b)\mu_i^q[k]}
    {\epsilon^2 a^2 e^{-b} +a\bar{a} (\epsilon^2+1-\frac{1}{2}\mu_i^q[k]\sigma_r^2) +\bar{a}^2e^b }$ & (\text{A7})\\
    $\overleftarrow{\xi}^r_i[k]=\Re\{\hat{q}_i[k]\}-\frac{1}{2} \overleftarrow{\psi}_i[k] 
    \frac{-ae^{-b}\sigma_r}{ae^{-b}+\bar{a}} $ & (\text{A8})\\
    $\overleftarrow{\xi}^i_i[k]=\Im\{\hat{q}_i[k]\}-\frac{1}{2} \overleftarrow{\psi}_i[k] 
    \frac{-ae^{-b}\sigma_i}{ae^{-b}+\bar{a}} $ & (\text{A9})\\
    $\overleftarrow{\xi}_i[k] = \overleftarrow{\xi}^r_i[k] + j\overleftarrow{\xi}^i_i[k]$ & (\text{A10})\\
    \hline
  \end{tabular}
\end{table}

\begin{figure*}[!h]
  \centering
  \includegraphics[width=5in]{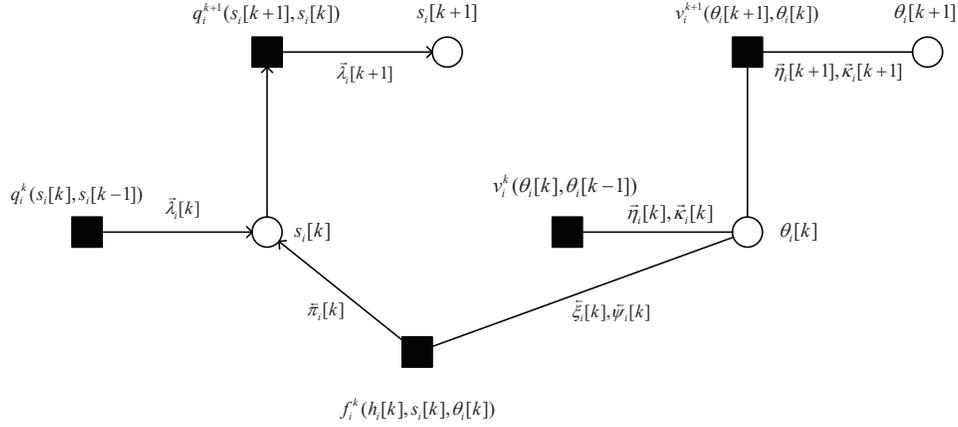}
  \caption{factor graph for \emph{across} stage}
  \label{across}
\end{figure*}
Following the \emph{out} stage, we have the \emph{across} stage as shown in Fig \ref{across}. Here, the soft information of the channel concerning 
joint channel estimation and equalization of one frame is passed to the next frame, enabling us to 
utilize information across multiple data frames. The messages from $q^{k+1}$ to $s_i[k+1]$ and from 
$v_i^{k+1}$ to $\theta_i[k+1]$ are updated and used to provide a more confident local prior for the next 
frame. These two messages can be computed using SPA directly.\par
To update $\mu_{q^{k+1}_i\rightarrow s_i[k+1]} $, $\mu_{s_i[k] \rightarrow q^{k+1}_i}$ is first computed as
\begin{equation}
  \begin{aligned}
    \mu_{s_i[k] \rightarrow q_i^{k+1} }&\propto\mu_{f^k_i \rightarrow s_i[k]}\mu_{q^k_i \rightarrow s_i[k] }\\
    &= (1-\overleftarrow{\pi}_i[k])(1-\overrightarrow{\lambda}_i[k])(1-s_i[k])\\
    &+\overleftarrow{\pi}_i[k]\overrightarrow{\lambda}_i[k]s_i[k],
  \end{aligned}
\end{equation}
and the final result is
\begin{equation}
  \begin{aligned}
    &\mu_{q^{k+1}_i \rightarrow s_i[k+1]}\propto \sum_{s_i[k]} p(s_i[k+1]|s_i[k]) \mu_{s_i[k] \rightarrow q^{k+1}_i}\\
    &=p(s_i[k+1]|s_i[k]=0)(1-\overleftarrow{\pi}_i[k])(1-\overrightarrow{\lambda}_i[k])\\
    &+p(s_i[k+1]|s_i[k]=1)\overleftarrow{\pi}_i[k]\overrightarrow{\lambda}_i[k]\\
     &=(1-\overrightarrow{\lambda}_i[k+1])(1-s_i[k+1])+\overrightarrow{\lambda}_i[k+1]s_i[k+1],
  \end{aligned}
\end{equation}
where
\begin{equation}\label{across lambda}
  \overrightarrow{\lambda}_i[k+1]=\frac{p_{11}\overleftarrow{\pi}_i[k]\overrightarrow{\lambda}_i[k]
  +p_{01}(1-\overleftarrow{\pi}_i[k])(1-\overrightarrow{\lambda}_i[k])}
  {(1-\overleftarrow{\pi}_i[k])(1-\overrightarrow{\lambda}_i[k])+\overleftarrow{\pi}_i[k]\overrightarrow{\lambda}_i[k]}.
\end{equation}
\par
$\mu_{v_i^{k+1} \rightarrow \theta_i[k+1]}$ can be derived using a similar approach. Specifically, 
\begin{equation}
  \begin{aligned}
    \mu_{\theta_i[k] \rightarrow v^{k+1}_i}&=\mu_{v_i^k \rightarrow \theta_i[k]}
    \mu_{f^k_i \rightarrow \theta_i[k]}\\
    &=\mathcal{CN}(\theta_i[k]|c, C),
  \end{aligned}
\end{equation}
where $C=\frac{\overrightarrow{\kappa}_i[k]\overleftarrow{\psi}_i[k]}{\overrightarrow{\kappa}_i[k]+\overleftarrow{\psi}_i[k]}$
, $c=\frac{\overrightarrow{\eta}_i[k]\overleftarrow{\psi}_i[k] + 
\overleftarrow{\xi}_i[k]\overrightarrow{\kappa}_i[k]}{\overrightarrow{\kappa}_i[k]+\overleftarrow{\psi}_i[k]}$ and 
\begin{equation}
  \begin{aligned}
    &\mu_{v_i^{k+1}\rightarrow \theta_i[k+1]}=\int p(\theta_i[k+1]|\theta_i[k])\mu_{\theta_i[k] \rightarrow v^{k+1}_i}d\theta_i[k]\\
    &=\int \frac{1}{1-\varrho} \mathcal{CN}\left(\theta_i[k]|\frac{\theta_i[k+1]-\zeta}{1-\varrho}+
    \zeta, \frac{\varrho^2 \rho}{(1-\varrho)^2}\right)\\
    &\times\mathcal{CN}(\theta_i[k]|c, C)d\theta_i[k]\\
    &=\frac{1}{1-\varrho}\mathcal{CN}\left(
      0|\frac{\theta_i[k+1]-\zeta}{1-\varrho}+\zeta-c, \frac{\varrho^2\rho}{(1-\varrho)^2}+C\right)\\
    &=\mathcal{CN}(\theta_i[k+1]|\overrightarrow{\eta}_i[k+1],\overrightarrow{\kappa}_i[k+1]),
  \end{aligned}
\end{equation}
with
\begin{equation}\label{across eta}
  \overrightarrow{\eta}_i[k+1]=(1-\varrho)c+\varrho\zeta
\end{equation}
\begin{equation}\label{across kappa}
  \overrightarrow{\kappa}_i[k+1] = (1-\varrho)^2C+\varrho^2 \rho.
\end{equation}
\par
The DCS-BiGAMP joint channel estimation and equalization algorithm is now complete. 
The output \emph{a posteriori} coefficients of the channel are then used to update the \emph{a priori} coefficients through the 
EM algorithm, and the output extrinsic coefficients of symbols are used to generate extrinsic LLRs.
\subsection{Extrinsic Log-likelihood Rate}
AMP-based estimation algorithms have a significant advantage that they can provide 
extrinsic information during the iteration process. In contrast, traditional algorithms like LMMSE \cite{tuchler2002minimum} \cite{tuchler2011turbo}, 
LMS and RLS\cite{947038} require modification in order to compute extrinsic LLRs and take into account the 
\emph{a priori} LLRs. As shown in Section \ref{twoA} substituting the Gaussian \emph{a priori} distribution with a 
discrete distribution allows for the natural incorporation of a priori LLRs. 
Additionally, computing extrinsic LLRs with the output of DCS-BiGAMP is also straightforward, which is presented as follows.\par
In the \emph{within} stage of DCS-BiGAMP, we have already computed the extrinsic expectation $\hat{r}_j[k]$ and variance 
$\mu^r_j[k]$, which is exactly what we need. The extrinsic LLR of the $q^{th}$ bit belonging to symbol $x_j[k]$ 
can be expressed as \footnote{\label{fn1}We omit the parameter $k$ here, which denotes the index of data frame, to simplify the expression.}
\begin{equation}\label{extllr}
  \begin{aligned}
    &L_{\text{ext}}(c_j,q)\\
    &=\ln \frac
    {\sum_{\alpha_n \in \mathcal{A}_q^0}
    \exp\left(\frac{\left\lvert \alpha_n-\hat{r}_j \right\rvert^2 }{\mu_j^r}\right)
    \prod_{q' \backslash q}p(c_{j,q'}=\chi_{n,q'})}
    {\sum_{\alpha_n \in \mathcal{A}_q^1}
    \exp\left(\frac{\left\lvert \alpha_n-\hat{r}_j \right\rvert^2 }{\mu_j^r}\right)
    \prod_{q' \backslash q}p(c_{j,q'}=\chi_{n,q'})}\\
    &=\ln \frac
    {\sum_{\alpha_n \in \mathcal{A}_q^0}
    \exp\left(\frac{\left\lvert \alpha_n-\hat{r}_j \right\rvert^2 }{\mu_j^r}
    \sum_{q'\backslash q}(-1)^{\chi_{n,q'}}\frac{L_{\text{apr}}(c_{j,q'})}{2}\right)}
    {\sum_{\alpha_n \in \mathcal{A}_q^1}
    \exp\left(\frac{\left\lvert \alpha_n-\hat{r}_j \right\rvert^2 }{\mu_j^r}
    \sum_{q'\backslash q}(-1)^{\chi_{n,q'}}\frac{L_{\text{apr}}(c_{j,q'})}{2}\right)},
  \end{aligned}
\end{equation}
where $\mathcal{A}_q^c$ represents the subset of $\mathcal{A}$ consisting of the symbol for which the $q^{th}$ bit is $c\in\{0, 1\}$.
\subsection{Updating the a priori Coefficient of Channel}
The a priori of symbols is updated by the decoder, and we use the expectation-maximization algorithm to
update the a priori coefficients $\mathbf{q}=[p_{01}, \lambda, \zeta, \varrho , \rho]$ on the channel side. Details are summarized in 
Table \ref{tab:EM}.\par

\begin{table}[!h]
  \caption{coefficients updated by EM}\label{tab:EM}
  \centering
  \begin{tabular}{|l r|}
    \hline
    $\lambda^{\text{new}}=\frac{1}{L}\sum_{i=1}^L\frac
    {\overleftarrow{\pi}_i[1]\lambda\overleftarrow{\lambda}_i[1]}
    {(1-\overleftarrow{\pi}_i[1])(1-\lambda)(1-\overleftarrow{\lambda}_i[1])}$ & (\text{B1})\\
    $p_{01}^{\text{new}}=\frac
    {\sum_{k=2}^{K}\sum_{i=1}^{L} E[s_i[k-1]|\bar{\mathbf{y}}] - E[s_i[k-1]s_i[k]|\bar{\mathbf{y}}] }
    {\sum_{k=2}^K\sum_{i=1}^LE[s_i[k-1]|\bar{\mathbf{y}}]}$ & (\text{B2})\\
    $\rho^{\text{new}}=\frac{1}{L(K-1)}\sum_{k=2}^K\sum_{i=1}^L
    \tilde{v}_i[k]+\left\lvert \tilde{\mu}_i[k] \right\rvert^2 + \left\lvert \zeta \right\rvert^2
    $ & \\
    $-2\Re\{ \tilde{\mu}_i[k]^* \zeta^k  \}$ & (\text{B3})\\
    $\zeta^{\text{new}}=\frac
    {\sum_{k=2}^K \sum_{i=1}^L \frac{1}{\varrho\rho} (\tilde{\mu}_i[k]-(1-\varrho)\tilde{\mu}_i[k-1])
    +\sum_{i=1}^L  \tilde{\mu}_i[1]/\sigma^2}
    {L((K-1)/\rho +1/\sigma^2)}$ & (\text{B4})\\
    $\varrho^{new} = \frac{1}{4 L (K-1)}(b-\sqrt{b^2+8L(K-1)c})$ & (\text{B5})\\
    $\tilde{v}_i[k] \triangleq V[\theta_i[k]|\bar{\mathbf{y}}]$ & (\text{B6})\\
    $\tilde{\mu}_i[k]\triangleq E[\theta_i[k]|\bar{\mathbf{y}}]$ & (\text{B7}) \\
    $b\triangleq \frac{2}{\rho} \sum_{k=2}^K\sum_{i=1}^L \Re\{E[\theta_i[k]^*\theta_i[k-1]|\bar{\mathbf{y}}]\}$ & \\
    $-\Re\{ (\tilde{\mu}_i[k] - \tilde{\mu}_i[k-1])^* \zeta  \}-\tilde{v}_i[k-1]-\left\lvert \tilde{\mu}_i[k-1] \right\rvert^2$
    & (\text{B8})\\
    $c \triangleq \frac{2}{\rho} \sum_{k=2}^K \sum_{i=1}^L \tilde{v}_i[k]+
    \left\lvert \tilde{\mu}_i[k] \right\rvert^2 + \tilde{v}_i[k-1] + \left\lvert \tilde{\mu}_i[k-1] \right\rvert^2
    $ & \\
    $-2\Re\{ E[\theta_i[k]^* \theta_i[k-1] | \bar{\mathbf{y}}] \}$ & (\text{B9})\\
    \hline
  \end{tabular}
\end{table}
Herein, we complete the entire description of the DCS-BiGAMP-based joint channel estimation and equalization/decoding algorithm. 
We name it DCS-JCED and summarize it in Algorithm~\ref{alg:dcs-jced}.

\begin{algorithm*}
  \caption{DCS-JCED}\label{alg:dcs-jced}
  \begin{algorithmic}
    \STATE\textbf{Input:} bit \emph{a priori} $L_{\text{apr}}$, 
    channel \emph{a priori} coefficients $\mathbf{q}=[p_{01}, \lambda, \zeta, \varrho , \rho]$, 
    AWGN noise variance $\sigma_w$, number of forward propagation $T_{\text{fp}}$, number of backward propagation 
    $T_{\text{bp}}$, number of inner BiGAMP iteration $T_{\text{inner}}$, number of turbo iteration $T_{\text{turbo}}$.
    \STATE\textbf{Initialization:}
    $\{ \hat{r}_j[k] \}_{j=1}^{N_p}=x_{p,j}$, 
    $\{ \hat{\mu}^r_j[k] \}_{j=1}^{N_p}=\epsilon$, $\{ \hat{r}_j[k] \}_{j=N_p+1}^{N_p+N_{\text{cd}}+N_g}=0$, $\{ \hat{\mu}^r_j[k] \}_{j=N_p+1}^{N_p+N_{\text{cd}}}=1$, \\
    $\{ \hat{\mu}_j^r[k] \}_{j=N_p+N_{\text{cd}}+1}^{N_p+N_{\text{cd}}+N_g}=\epsilon$, $\{ \hat{q}_i[k] = q^{\text{init}}_i[k]\}_{i=1}^{L}$, $\{ \hat{\mu}^q =(\mu^{q}[k])^{\text{init}}\}_{i=1}^L$, 
    $\{ \hat{s}_m \}_{m=1}^{M}=0$, $\epsilon \to 0$.
    \STATE \textbf{for} $t_{\text{turbo}}=1,\dots, T_{\text{turbo}}$ \textbf{do}
    \STATE initialize $t_{\text{fp}}=0$, $t_{\text{bp}}=0$.
    \STATE \textbf{if} $t_{\text{fp}} < T_{\text{fp}}$ \textbf{do}
    \STATE $t_{\text{fp}} = t_{\text{fp}} + 1$
    \STATE \textbf{for} $k=1,\dots, K$ \textbf{do forward propagation}
    \STATE \hspace{0.5cm} \textbf{into stage} $\forall i$  calculate local prior $\overrightarrow{\pi}_i[k]$, 
    $\overrightarrow{\xi}_i[k]$, $\overrightarrow{\psi}_i[k]$ by (\ref{into pi}), (\ref{into xi}), (\ref{into psi})
    \STATE \hspace{0.5cm} \textbf{within stage}
    \STATE \hspace{0.5cm} \textbf{for} $t_{\text{inner}}=1, 2, \dots, T_{\text{inner}}$
    \STATE \hspace{1.0cm} \textbf{step \uppercase\expandafter{\romannumeral1}} $\forall i$ update \emph{a posteriori} coefficients of $\hat{h}_i[k]$, 
    $\mu^h_i[k]$ by (\ref{within hhat}), (\ref{within hvar})
    \STATE \hspace{1.0cm} $j \in [N_p+1, N_p+N_{\text{cd}}]$ update \emph{a posteriori} coefficients of $\hat{x}_j[k]$, 
    $\mu^x_j[k]$ by (\ref{within xhat}), (\ref{within xvar})
    \STATE \hspace{1.0cm} \textbf{step \uppercase\expandafter{\romannumeral2}} $\forall m$ update the conditional coefficients of $\mathbf{z}[k]$, i.e. $\hat{p}_m[k]$, $v^p_m[k]$ by (\ref{within phat}), (\ref{within pvar})
    \STATE \hspace{1.0cm} \textbf{step \uppercase\expandafter{\romannumeral3}} $\forall m$ update the \emph{a posteriori} coefficients of $\mathbf{z}[k]$, i.e. 
    $\hat{z}_m[k]$, $\mu^z_m[k]$ by (\ref{within zhat}), (\ref{within zvar})
    \STATE \hspace{1.0cm} \textbf{step \uppercase\expandafter{\romannumeral4}} $\forall m$ update the Taylor coefficients $\hat{s}_m[k]$, $\mu^s_m[k]$
    by (\ref{within shat}), (\ref{within svar})
    \STATE \hspace{1.0cm} \textbf{step \uppercase\expandafter{\romannumeral5}} $\forall i$ update the extrinsic coefficients of $\hat{q}_i[k]$, 
    $\mu^q_i[k]$ by (\ref{within qhat}), (\ref{within qvar})
    \STATE \hspace{1.0cm} $j \in [N_p+1, N_p+N_{\text{cd}}]$ update the extrinsic coefficients of 
    $\hat{r}[k]$, $\mu^r_i[k]$ by (\ref{within rhat}), (\ref{within rvar})
    \STATE \hspace{0.5cm} \textbf{endfor}
    \STATE \hspace{0.5cm} \textbf{out stage} $\forall i$ calculate $\overleftarrow{\pi}_i[k]$, $\overleftarrow{\xi}_i[k]$, 
    $\overleftarrow{\psi}_i[k]$ by (\ref{out pi}), Table \ref{tab:taylor}
    \STATE \hspace{0.5cm} \textbf{forward across stage} $\forall i$ calculate $\overrightarrow{\lambda}_i[k+1]$, $\overrightarrow{\eta}_i[k+1]$, 
    $\overrightarrow{\kappa}_i[k+1]$ by (\ref{across lambda}), (\ref{across eta}), (\ref{across kappa})
    \STATE \textbf{end for}
    \STATE \textbf{end if}
    \STATE \textbf{if} $ t_{\text{bf}} < T_{\text{bf}}$ \textbf{do}
    \STATE $t_{\text{bf}} = t_{\text{bf}} + 1$
    \STATE \textbf{backward propagation}
    \STATE \textbf{end if}
    \STATE \textbf{EM tuning}
    \STATE update $\mathbf{q}=[p_{01}, \lambda, \zeta, \varrho , \rho]$ by Table \ref{tab:EM}
    \STATE \textbf{decoding}
    \STATE calculate $L_{\text{ext}}$ by (\ref{extllr}) and send into decoder to get 
    $L_{\text{apr}}$
    \STATE \textbf{end for}
  \end{algorithmic}
\end{algorithm*}

\section{numerical simulation}
We present the numerical results of DCS-JCED in comparison with the original JCED and classic CE-based LMMSE 
turbo equalizer. We evaluate the performance from multiple perspectives, including BER, 
NMSE of the channel under different signal-to-noise ratios ($E_b/N_0$), pilot-to-data 
rate $N_p/(N_p+N_{\text{cd}})$, time of forward propagation $T_{\text{fp}}$ and backward propagation $T_{\text{pb}}$, the number 
of data frames $K$ demodulated at the same time.
\subsection{Simulation Setup}
\begin{figure}[!t]
  \centering
  \includegraphics[width=3.25in]{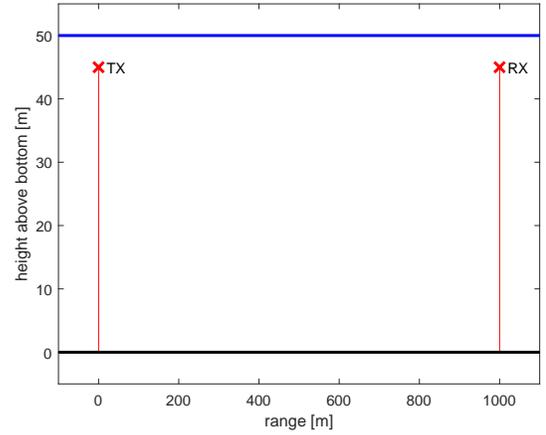}
  \caption{UWA simulation environment.}
  \label{sim_env}
\end{figure}

\begin{figure}[!t]
  \centering
  \includegraphics[width=3.25in]{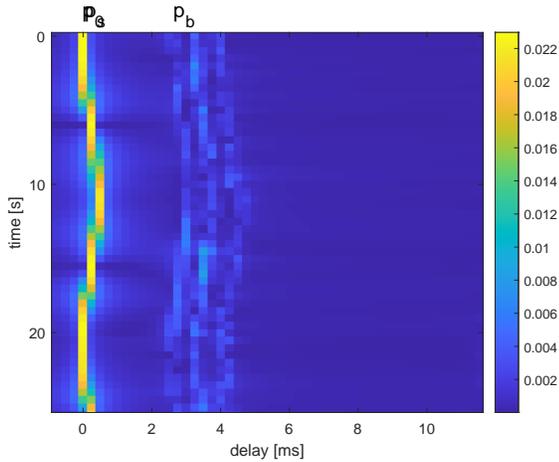}
  \caption{Simulated channel impulse responses.}
  \label{sim_channel}
\end{figure}

To improve simulation realism, we use the UWA model proposed in \cite{2013Statistical} 
and its MATLAB toolbox\cite{qarabaqi2013acoustic}. This model has been widely accepted and verified by a 
large amount of sea trial data obtained by Surface Processes and
Acoustic Communications Experiment (SPACE), Mobile Acoustic Communications
Experiment (MAC) and Kauai AComms Multidisciplinary University Research Initiative (KAM). 
Here, we set the depth of the sea to 50~m and place the transmitter and receiver at a 
depth of 5~m with a distance of 1000~m, as shown in Fig~\ref{sim_env}. 
The range of the surface height variation, transmitter height variation, and receiver height variation 
is set between -1.25~m and 1.25~m, while the range of channel distance variation is set between -20~m and 
20~m. Under these conditions, we generated 200 channel, and an example of the channel is provided in Fig.~\ref{sim_channel}.\par

\begin{figure}[!t]
  \centering
  \includegraphics[width=3.25in]{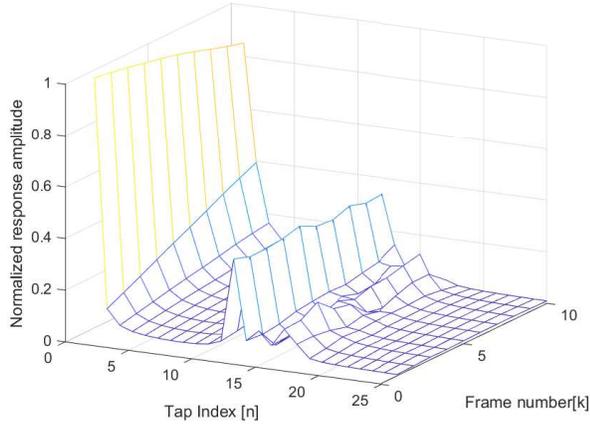}
  \caption{Absolute-value amplitude of example channel for 10 data frames in transmission simulation.}
  \label{sim_channel_exp}
\end{figure}

In the simulation, we set the transmission symbol rate to 4~kbit/s, which leads to a tap delay resolution of 0.25~ms, 
as in Fig.~\ref{sim_channel}. The dominant components of the 
channel delay taps are concentrated within the first 6 milliseconds. Therefore, we take the first 25 taps as the true 
channel in our simulation. The frame structure is as follows: the length of guard sequence $N_g$ is set to 25, 
the pilot is an M-sequence modulated by phase shift keying (PSK), and the length of pilot $N_p$ is selected from $31$ or $63$ in 
different simulations. $N_b=130$ data bits are coded at a rate of $1/2$ using LDPC code and then mapped to 
quad-phase shift key (QPSK), 
resulting in $N_{\text{cd}}=130$ data symbols per data frame. In this case, the total data frame length 
is selected from $N_c=$186 or 218, such that the maximum time duration of a data frame is 0.0545~s. Given a  
time resolution of $0.05$~s, we take one row of the simulated channel through which a data frame passes. 
For a transmission containing 10 data frames, one example channel is shown in Fig~\ref{sim_channel_exp}.

\subsection{Performance analysis}
\begin{figure}[!t]
  \centering
  \includegraphics[width=3.25in]{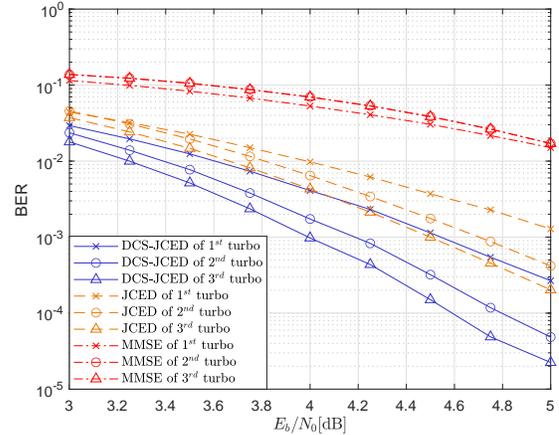}
  \caption{BER versus $E_b/N_0$ for $K = 10$, $N_p = 63$, $T_{\text{fp}}=T_{\text{bp}}=2$.}
  \label{pilot_63_package_10_ber}
\end{figure}

\begin{figure}[!t]
  \centering
  \includegraphics[width=3.25in]{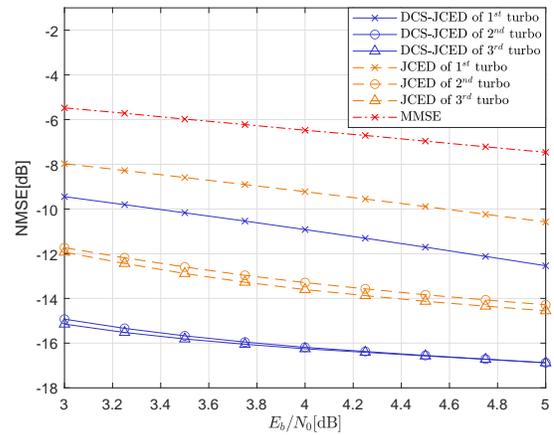}
  \caption{Estmiated channel NMSE versus $E_b/N_0$ for $K = 10$, $N_p = 63$, $T_{\text{fp}}=T_{\text{bp}}=2$.}
  \label{pilot_63_package_10_nmse}
\end{figure}

We first analyze the relationship between BER, estimated channel NMSE and bit SNR $E_b/N_0$. 
We set $N_p=63$ and transmit 10 data frames at a time, with the forward propagation $T_{\text{fp}}$ and $T_{\text{bf}}$ noth set 
to 2. We simulate 200 channels and run 200 simulations for each channel. 
The MMSE turbo equalizer uses the LMMSE algorithm to estimate 
the channel, while both JCED and DCS-JCED use a Gaussian random sequence to initialize the 
channel. 
For fairness, we use the \emph{a posteriori} channel estimates of the $k^{\text{th}}$ data frame to 
initialize the channel of the $(k+1)^{\text{th}}$ data frame when performing JCED on the $(k+1)^{\text{th}}$ data frame. 
Furthermore, the number of iterations $T_{\text{inner}}$ is 100 for JCED and 25 for DCS-JCED. 
For all methods, we perform turbo iteration 3 times. The lengths of the non-casual and causal parts 
of the filter $N_1$ and $N_2$ in the MMSE turbo equalizer \cite{tuchler2002minimum} are set to 15 and 20, respectively. 
The initial parameters of the Bernoulli Gaussian distribution for DCS-JCED and JCED 
are set to $\lambda=0.2$, $\zeta = 0$, $\rho = 1$. For parameters that 
characterize the time-varying feature of the channel, we initialize them by 
setting $p_{01} = 0.01$ and $\varrho  = 0.005$. 
We set the breakout condition of the iteration 
$\Vert\hat{\boldsymbol{z}}^t[k]- \hat{\boldsymbol{z}}^{t-1}[k] \Vert_2/
\Vert\hat{\boldsymbol{z}}^{t-1}[k]\Vert_2 < 10^{-4}$, where $\hat{\boldsymbol{z}}^t[k]$ 
is the estimation of $\mathbf{z}[k]$ in the current iteration, and $\hat{\boldsymbol{z}}^{t-1}[k]$ is 
the estimation in the last iteration. 
The BER curves are shown in Fig.~\ref{pilot_63_package_10_ber}, and the NMSE curves 
of the estimated channel are shown in Fig.~\ref{pilot_63_package_10_nmse}. 
Due to the low SNR and short pilot length, the NMSE of the LMMSE channel 
estimator is relatively high, resulting in poor BER performance in equalization. 
At this point, the turbo equalization operation impacts negatively since the results 
of the equalizer are no longer reliable with such a high BER. 
The original JCED performs well in this scenario, since it can jointly utilize the soft information of both 
the channel and symbols. Turbo equalization not only reduces the BER but also significantly reduces 
the NMSE of the channel estimates between one and two turbo iterations. 
Building on this groundwork, DCS-JCED utilizes the information of multiple data frames by passing messages 
across them, thus achieving the best BER and NMSE performance. Note that DCS-JCED reduces the NMSE of 
channel estimates more rapidly between one and two turbo iterations than the 
original JCED. \par


\begin{figure}[!t]
  \centering
  \includegraphics[width=3.25in]{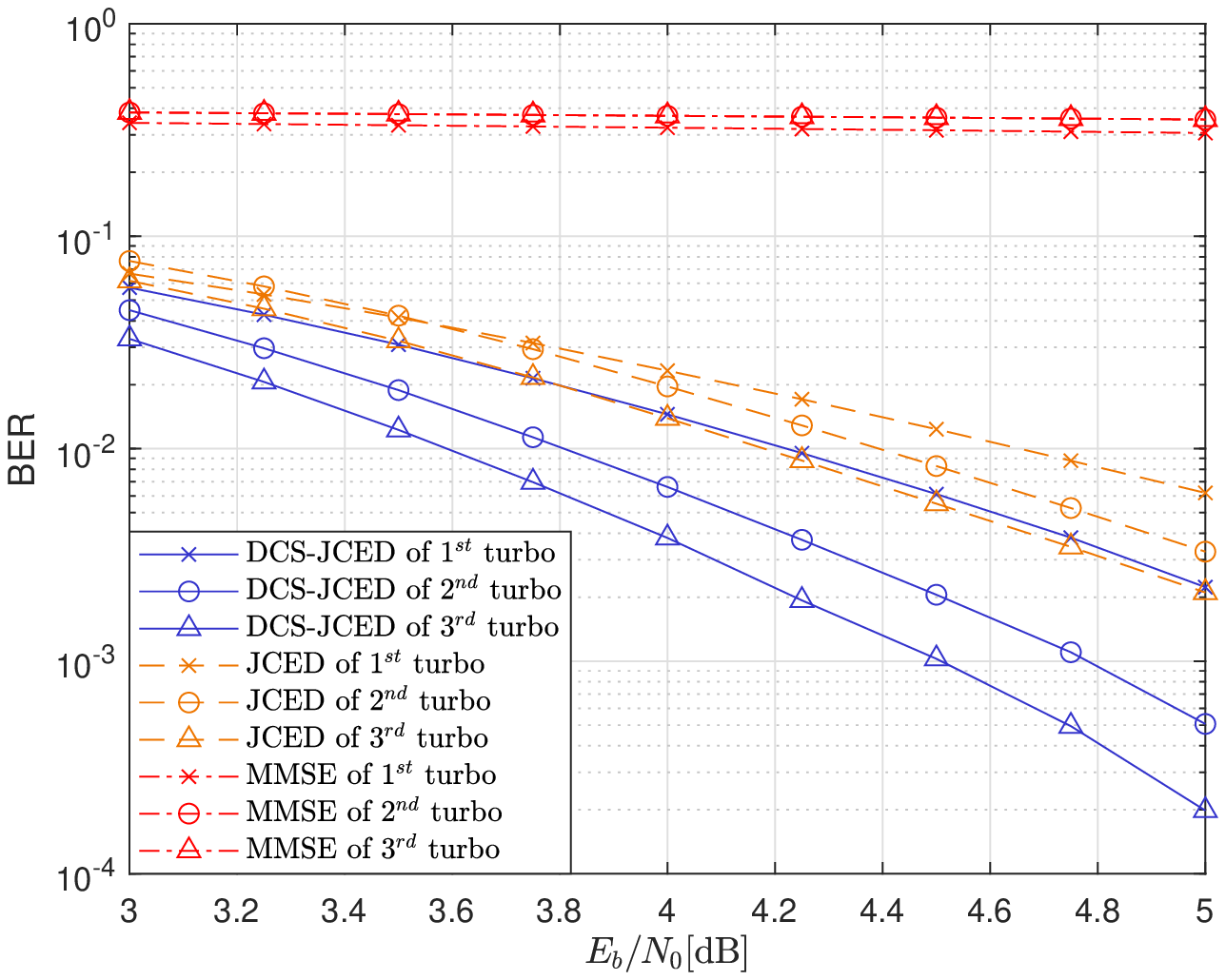}
  \caption{BER versus $E_b/N_0$ for $K = 10$, $N_p = 31$, $T_{\text{fp}}=T_{\text{bp}}=2$.}
  \label{pilot_31_package_10_ber}
\end{figure}

\begin{figure}[!t]
  \centering
  \includegraphics[width=3.25in]{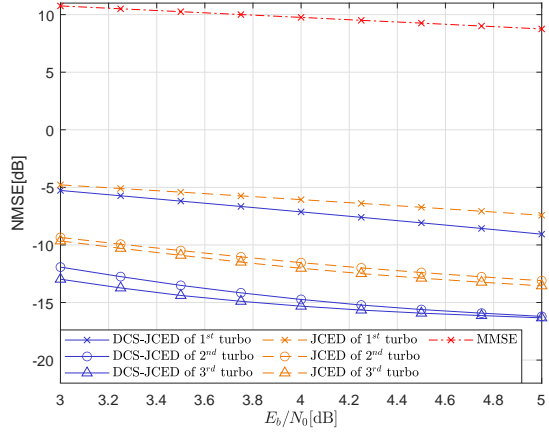}
  \caption{Estmiated channel NMSE versus $E_b/N_0$ for $K = 10$, $N_p = 31$, $T_{\text{fp}}=T_{\text{bp}}=2$.}
  \label{pilot_31_package_10_nmse}
\end{figure}

After changing the pilot number, while keeping other parameters the same, to 31, 
the BER and estimated channel NMSE curves in Fig.~\ref{pilot_31_package_10_ber} and \ref{pilot_31_package_10_nmse} show a loss of performance across all methods.  
The MMSE equalizer fails completely since the pilot length is too short for LMMSE channel estimator 
to obtain a reliable result. The NMSE of the LMMSE channel estimator is much higher than that of the 
original JCED and the proposed DCS-JCED. Although both the JCED and DCS-JCED suffer a performance loss, their NMSE 
remains at a low level, and their BER results are an order of magnitude worse compared to 
the results using $N_p=63$. It should be noted that to achieve the same NMSE performance as 
that with $N_p=63$, the LMMSE channel estimator requires an $E_b/N_0$ of approximately $20$dB. 
At such an $E_b/N_0$ level, the JCED and DCS-JCED already achieve error-free transmission. When 
comparing the results of JCED and DCS-JCED, we observe that DCS-JCED again has the best 
BER and NMSE performance. \par

\begin{figure}[!t]
  \centering
  \includegraphics[width=3.25in]{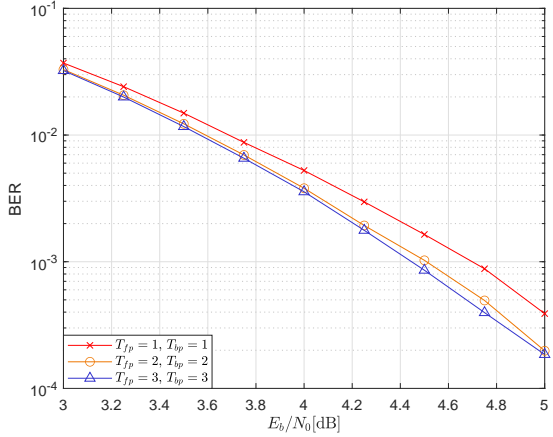}
  \caption{BER versus $E_b/N_0$ of $3^{rd}$ turbo for $K=10$, $N_p=31$
  and different $T_{\text{fp}}$, $T_{\text{bp}}$ in DCS-JCED.}
  \label{pilot_31_fb_123_ber}
\end{figure}

\begin{figure}[!t]
  \centering
  \includegraphics[width=3.25in]{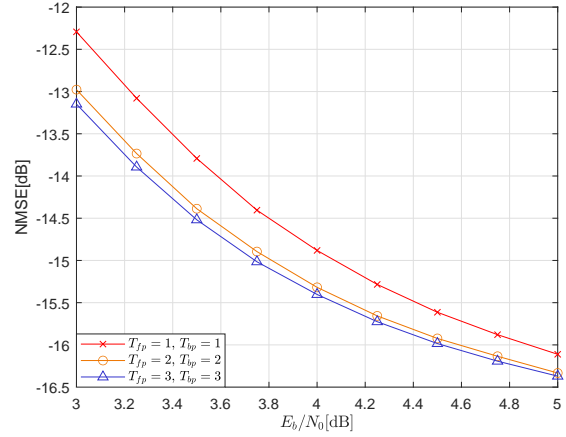}
  \caption{Estmiated channel NMSE versus $E_b/N_0$ of $3^{\text{rd}}$ turbo iteration for $K=10$, $N_p=31$
  and different $T_{\text{fp}}$, $T_{\text{bp}}$ in DCS-JCED.}
  \label{pilot_31_fb_123_nmse}
\end{figure}

Since the proposed DCS-JCED uses both forward and backward propagation to pass 
messages between every data frame, we evaluate its performance with respect to the propagation time. 
We conduct simulations by setting 1, 2 and 3 forward and backward propagations, 
with $N_p=31$, $K=10$, and $T_{\text{inner}}=50$. 
BER curves versus $E_b/N_0$ after 3 turbo iterations are shown in Fig. \ref{pilot_31_fb_123_ber}. 
It is evident that the performance of DCS-JCED gradually improves with the increased 
$T_{\text{fp}}$ and $T_{\text{bp}}$. It can be interpreted that each data frame receives information from 
other data frames through \emph{across}-stage message passing, and with an 
increase in the propagation times, messages passed among them becomes more reliable. 
However, this benefit will gradually decreases as the number of propagation increases, 
since more propagations will not provide extra information for \emph{within}-stage processing. 
Fig.~\ref{pilot_31_fb_123_ber} illustrates that, the performance gain 
between $T_{\text{fp}}=T_{\text{bp}}=1$ and $T_{\text{fp}}=T_{\text{bp}}=2$ is much higher than that between 
$T_{\text{fp}}=T_{\text{bp}}=2$ and $T_{\text{fp}}=T_{\text{bp}}=3$. This phenomenon  can also be observed in the 
estimated channel NMSE curves versus $E_b/N_0$ as shown in Fig.~\ref{pilot_31_fb_123_nmse}. 
It is reasonable to predict that both the BER and NMSE curves versus $E_b/N_0$ will converge to 
a fixed curve with more forward and backward propagations.\par

\begin{figure}[!t]
  \centering
  \includegraphics[width=3.25in]{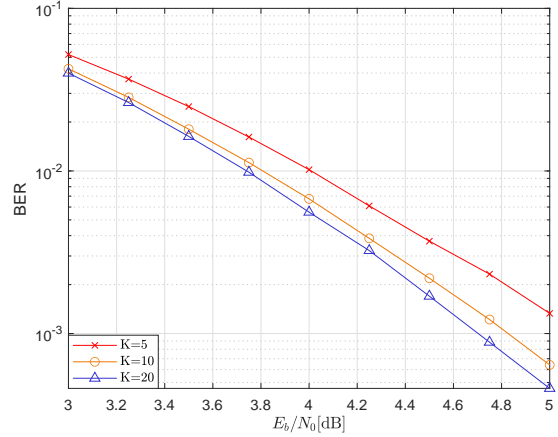}
  \caption{BER versus $E_b/N_0$ of $3^{\text{rd}}$ turbo iteration for $N_p=31$, $T_{\text{fp}}=1$, 
  $T_{\text{bp}}=1$ and different $K$ in DCS-JCED.}
  \label{pilot_31_pkt5_10_20_ber}
\end{figure}

\begin{figure}[!t]
  \centering
  \includegraphics[width=3.25in]{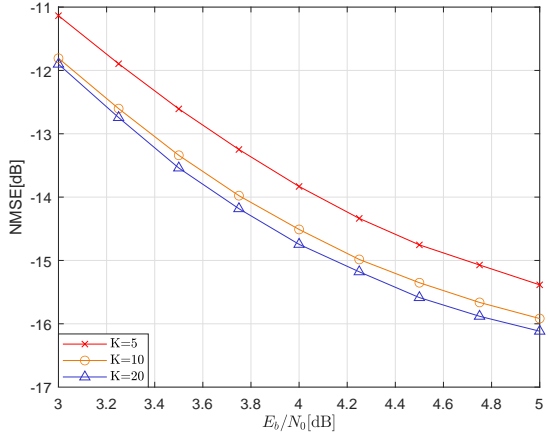}
  \caption{Estmiated channel NMSE versus $E_b/N_0$ of $3^{\text{rd}}$ turbo for $N_p=31$, $T_{\text{fp}}=1$, 
  $T_{\text{bp}}=1$ and different $K$ in DCS-JCED.}
  \label{pilot_31_pkt5_10_20_nmse}
\end{figure}

The proposed DCS-JCED algorithm has the main feature of utilizing information 
across multiple data frames. Consequently, we analyze the performance of DCS-JCED with varying numbers of 
data frames in transmission. If we transmit multiple data frames, we may divide them into several groups 
and perform DCS-JCED on each group at the receiver. 
In the numerical simulation, we set the total number of transmitted data frames to 20, and at the 
receiver, we divide them into 4, 2 and 1 group(s) and use DCS-JCED to demodulate them, i.e. 
demodulating 5 data frames for 4 times, 10 data frames for 2 times and 20 data frames at once. 
The results of the BER and estimated channel NMSE curves versus $E_b/N_0$ are shown in 
Fig.~\ref{pilot_31_pkt5_10_20_ber} and \ref{pilot_31_pkt5_10_20_nmse} respectively. 
As expected, the performance of DCS-JCED improves as the number of data frames being demodulated at once, i.e. $K$, increases. 
This is because each data frame exchanges information with others, and therefore, 
the DCS-JCED algorithm can utilize more information as the number of data frames increases. 
However, the performance gain generated by this effect also tends to saturate 
with the increase of $K$. One can observe that both the BER and NMSE improvement 
between $K=20$ and $K=10$ is less than that between $K=10$ and $K=5$, 
which suggests that the information exchange between different data frames reaches a limit with the increasing number of data frames.

\section{Conclusions}
In this paper, we propose a joint channel estimation and symbol equalization/decoding equalizer 
named DCS-JCED with fully soft-in and soft-out functionality. 
Our equalizer is based on the DCS-BiGAMP algorithm which combines the strengths of 
existing DCS-AMP and BiGAMP algorithms. By using DCS-JCED, we can fully utilize the soft information 
of both the channel and symbols across multiple data frames, whereas traditional JCED algorithm based on 
BiGAMP only considers information within a single data frame. The proposed equalizer can also be performed in both the 
time and frequency domains. \par
Numerical simulations conducted on a time-varying UWA channel demonstrate that the proposed algorithm outperforms 
the classic CE-based LMMSE equalizer and the original JCED equalizer in both the BER and estimated 
channel NMSE performance. \par
The DCS-BiGAMP algorithm employed in the equalizer design is also applicable to 
other bilinear compressive sensing 
problems that have multiple measurements vectors, such as video denoising. 

{\appendix[Within Stage in Frequency Domain]

Here, we provide a frequency-domain version of the iteration in the \emph{within} stage. The 
computational procedure for the \emph{a posteriori} expectation and variance of $\mathbf{x}[k]$ and $\mathbf{h}[k]$ 
remains the same as that in the time domain.
The conditional expectation and variance of $\mathbf{z}[k]$ in the frequency domain, 
$\bar{\mathbf{z}}[k]=\mathbf{F}_M \mathbf{z}[k]/\sqrt{M}$, are
\begin{equation}
  \mathbf{v}^p[k]=\bar{\mathbf{v}}^p[k]+\left(
    \frac{1}{M} \sum_{j=1}^M \mu_j^x[k]
  \right)
  \left(
    \frac{1}{M} \sum_{l=1}^L \mu_l^h[k]
  \right)\mathbf{1}_M
\end{equation}
\begin{equation}
  \hat{\mathbf{p}}[k]=(\mathbf{F}_M \hat{\mathbf{x}}[k])\odot(\mathbf{F}_M^{1:L} \hat{\mathbf{h}}[k])
  -\hat{\mathbf{s}}[k]\odot \bar{\mathbf{v}}^p[k],
\end{equation}
where
\begin{equation}
  \begin{aligned}
    \bar{\mathbf{v}}^p[k]&=\frac{1}{M}\sum_{j=1}^{M}\mu_j^x[k] \left\lvert \mathbf{F}_M^{1:L} \hat{\mathbf{h}}[k] \right\rvert
  ^{\odot 2}\\
  &+\frac{1}{M}\sum_{l=1}^{L}\mu_l^h[k] \left\lvert \mathbf{F}_M \hat{\mathbf{x}}[k] \right\rvert
  ^{\odot 2}.
  \end{aligned}
\end{equation}
$\hat{\mathbf{s}}[k]$ and $\boldsymbol{\mu}^s[k]$ are the same as those in the time-domain version. To 
calculate the extrinsic parameters of $\mathbf{x}[k]$ and $\mathbf{h}[k]$, we have 
\begin{equation}
  \begin{aligned}
    \boldsymbol{\mu}^r[k] &= M/\left(
      (\boldsymbol{\mu}^s[k])^\mathsf{T} \left\lvert  \mathbf{F}_M^{1:L} \hat{\mathbf{h}}[k] \right\rvert^{\odot 2} 
  \right)\mathbf{1}_L\\
  &=\mu^r[k] \mathbf{1}_L
  \end{aligned}
\end{equation}
\begin{equation}
  \begin{aligned}
    \boldsymbol{\mu}^q[k] &= M/\left(
      (\boldsymbol{\mu}^s[k])^\mathsf{T} \left\lvert \mathbf{F}_M \hat{\mathbf{x}}[k] \right\rvert^{\odot 2} 
  \right)\mathbf{1}_M\\
  &=\mu^q[k] \mathbf{1}_M
  \end{aligned}
\end{equation}
and
\begin{equation}
  \begin{aligned}
    \hat{\mathbf{r}}[k]&=\hat{\mathbf{x}}[k]+\mu^r[k] \mathbf{F}_M^\mathsf{H}
  \left(
    (\mathbf{F}_M^{1:L} \hat{\mathbf{h}}[k])^{*}\odot \hat{\mathbf{s}}[k]
  \right) \\
  &- \mu^s[k] \mu^r[k] \mu^h[k] \hat{\mathbf{x}}[k]
  \end{aligned}
\end{equation}
\begin{equation}
  \begin{aligned}
    \hat{\mathbf{q}}[k]&=\hat{\mathbf{h}}[k]+\mu^q[k] {\mathbf{F}_M^{1:L}}^\mathsf{H}
  \left(
    (\mathbf{F}_M \hat{\mathbf{x}}[k])^{*}\odot \hat{\mathbf{s}}[k]
  \right) \\
  &- \mu^s[k] \mu^q[k] \mu^x[k] \hat{\mathbf{h}}[k],
  \end{aligned}
\end{equation}
where
\begin{equation}
  \mu^s[k] = \frac{1}{M}\sum_{m=1}^M \mu_m^s[k]
\end{equation}
\begin{equation}
  \mu_h[k] = \frac{1}{L}\sum_{l=1}^L \mu_l^h[k]
\end{equation}
\begin{equation}
  \mu^x[k] = \frac{1}{M}\sum_{j=1}^M \mu_j^x[k].
\end{equation}.
}\label{apdix}

\bibliography{IEEEabrv,myrefs}
\bibliographystyle{IEEEtran}

\vfill

\end{document}